\documentclass[12pt,a4paper]{article}

\usepackage{verbatim}
\usepackage[utf8]{inputenc}
\usepackage[T1]{fontenc}
\usepackage{lmodern}
\usepackage[margin=2.5cm]{geometry}
\usepackage{setspace}      
\usepackage{authblk}       
\usepackage{graphicx}
\usepackage{subfigure}
\usepackage{caption}
\usepackage{subcaption}
\usepackage{booktabs}
\usepackage{amsmath,amssymb}
\usepackage[numbers,sort&compress]{natbib}
\usepackage{hyperref}
\usepackage{xcolor}
\usepackage{tcolorbox}     

\title{A New Lens on the Sustainability of the AI Revolution }

\author{Pierluigi Contucci$^{1}$, Godwin Osabutey$^{1}$, and Filippo Zimmaro$^{1,2,*}$}

\affil{
$^{1}$Department of Mathematics, University of Bologna, Bologna, Italy\\
$^{2}$Department of Computer Science, University of Pisa, Pisa, Italy\\
*Correspondence to: \texttt{zimmarofilippo@gmail.com}\\
E-mail: \texttt{pierluigi.contucci@unibo.it}, \texttt{gosabutey@unimore.it}
}

\date{\today} 

\begin{document}
\maketitle

\begin{abstract}
We introduce the Economic Productivity of Energy (EPE), GDP generated per unit of energy consumed, as a quantitative lens to assess the sustainability of the Artificial Intelligence (AI) revolution. Historical evidence shows that the first industrial revolution, pre-scientific in the sense that technological adoption preceded scientific understanding, initially disrupted this ratio: EPE collapsed as profits outpaced efficiency, with poorly integrated technologies, and recovered only with the rise of scientific knowledge and societal adaptation. Later industrial revolutions, such as electrification and microelectronics, grounded in established scientific theory, did not exhibit comparable declines.
Today’s AI revolution, highly profitable yet energy-intensive, remains pre-scientific and may follow a similar trajectory in EPE. We combine this conceptual discussion with cross-country EPE data spanning the last three decades. 
We find that the advanced economies exhibit a consistent linear growth in EPE: those countries are the ones that contribute most to global GDP production and energy consumption, and are expected to be the most affected by the AI transition.
Therefore, we advocate for regular monitoring of EPE: transparent reporting of AI-related energy use and productivity-linked incentives can expose hidden energy costs and prevent efficiency-blind economic expansion. Embedding EPE within sustainability frameworks would help align technological innovation with energy productivity, a critical condition for sustainable growth.
\end{abstract}



\section{Introduction}

The relation between energy and economic growth is one of the defining issues of our time. Modern societies require amounts of energy hundreds of times larger than those of the pre--industrial era. This demand underpins global GDP but it also increases energy consumption, contributing to the exploitation of non-renewable sources and to greenhouse gas emission. Sustainability has become the threshold that growth can no longer ignore.

In this context we introduce the notion of \emph{Economic Productivity of Energy} (EPE). It is defined as the ratio
\[
\mathrm{EPE} = \frac{\text{GDP}}{\text{Energy}} \quad [\mathrm{US\$}/\mathrm{kWh}],
\] 
namely the amount of monetary value generated per unit of energy consumed. This quantity is the inverse of the more commonly used energy intensity~\cite{stern2018energy, stern2019energy}. Indeed, EPE has been considered a rough proxy for a country’s energy efficiency~\cite{Ang_2006}. The simple inversion changes completely the perspective: energy intensity quantifies how much energy is needed to obtain one unit of GDP, while EPE measures how much GDP can be produced from a fixed budget of energy. EPE highlights the productive capacity of societies under energy constraints and connects naturally to the concept of economic efficiency. Drawing an analogy with the thermal efficiency of a heat engine is instructive: similar ideas have been explored by studies that integrate thermodynamic concepts into economic analysis~\cite{ayres1998eco,burley2013economics,fujun2023metaheu}.
We believe this shift is important to reveal patterns otherwise hidden: for example, we find that (see Figure 2) EPE of advanced countries grows linearly over time, a behavior not observed when considering energy intensity.

From a historic analysis of the EPE trend, we observe that in the early stages of the first industrial revolution in England and Wales EPE collapsed~\cite{malanima2016energy}. Machines were profitable but energetically inefficient: economic output rose, but the ratio of GDP per unit of energy fell sharply. Only decades later, with more energy efficient machines and developed organization, did EPE recover.
When instead technology followed science, as with electricity in the second industrial revolution and quantum mechanics in the third, the trajectory of energy productivity was smoother, often with an increasing trend \cite{kanderinternational}: electrification, grounded in Maxwell’s theory, proved largely energy-saving \cite{mokyr1998secondIR, kander2014power}, while microelectronics, rooted in quantum semiconductor physics, delivered transistors and computing devices whose efficiency gains translated into sustained improvements in energy productivity \cite{brasenNBERtransistor, foster2023efficiency}.

Today we stand in the midst of the Artificial Intelligence revolution, running at full steam. AI is already reshaping entire sectors of the economy, as only a true industrial revolution can do. It is also extremely energy--hungry. Training large language models requires gigawatt-hours of energy, and inference at scale adds a continuous load. The analogy with the early phase of the first industrial revolution is immediate: a new technology adopted because it is profitable, without regard for its energetic cost~\cite{contucci2019intelligenza}. The current risk is that EPE may once again decline at the very moment when sustainability is at the center of the world's agenda.

At the same time, AI carries the potential to optimize the very processes that determine EPE: industrial production, logistics, energy grids, social organization. It is a paradoxical technology, able to threaten and to improve energy productivity at once. The outcome will depend on how it is adopted, regulated, integrated into society and, especially, on the understanding of the scientific principles at the root of this technology.

In this paper we argue that monitoring EPE is essential to understand the energetic dimension of the AI industrial revolution. GDP growth projections attributed to AI are impressive, but without reference to energy they are incomplete. The ratio of GDP to energy is the quantity that matters for sustainability. History shows that technological revolutions perturb it. The question is not whether AI will change EPE, but how and when.

We proceed with the following analysis. Section 2 reviews the historical trajectory of EPE during the first and other industrial revolutions, Section 2 discusses the recent trends of EPE across different macro classification groups of countries. Section 4 discusses the risks and opportunities for EPE in relation to AI as a pre-scientific engine, while Section 5 proposes EPE measurement agenda and policy implications. We conclude by proposing EPE as a critical indicator in the transition now underway.

\section{EPE across industrial revolutions (historical evidence)}

The long–term relation between energy use and wealth production can be described in terms of the Economic Productivity of Energy (EPE), defined as GDP produced per unit of energy consumed. This ratio provides a historical perspective on how societies have transformed energy flows into economic value. Evidence across several centuries shows that EPE does not follow a monotone trajectory. Instead, it exhibits discontinuities whenever a major technological revolution takes place.

From the sixteenth to the nineteenth century, European aggregate data indicate a persistent increase in per capita energy consumption~\cite{gentvilaite2014role,kander2014power}. The growth is not smooth: it accelerates after the introduction of new energy–using technologies and slows in periods of stagnation. The introduction of coal and steam, electricity and oil, and later information and communication technologies all coincide with distinct regime shifts. At each transition, the relation between GDP and energy changes, and the EPE responds accordingly. The first industrial revolution provides the clearest example.

The case of England and Wales between 1560 and 1900 has been studied in detail~\cite{malanima2016energy}. Before industrialization, EPE was about 0.21 US\$/kWh. With the introduction of steam engines and coal, this value nearly halved, dropping to 0.11 during the early phases of the revolution. In the following decades it recovered slightly, to around 0.13, but remained well below the pre–industrial level. The decline is unmistakable: economic output increased, but the energy required to produce it grew even faster. In the same period, Italy displayed a different pattern. There, where the industrial transition arrived later, EPE remained stable. This contrast shows that the fall in EPE is specific to the initial phase of industrial revolutions, when technology is profitable but inefficient.

In the first stages of a new technology, machines are adopted because they generate profit even if they consume large amounts of energy. Scientific understanding is incomplete, engineering practice is rudimentary, and regulation is absent. Private actors pursue immediate return, not long–term efficiency. Under these conditions GDP can rise while EPE falls. Only with subsequent technical improvements, social organization, and regulatory frameworks does the trend reverse. EPE begins to climb again when devices become more efficient, energy systems are better integrated, and production is reorganized around the new paradigm.
There are historical evidences \cite{kanderinternational, nielsen2018east}, although restricted to a group of Western countries, showing how the EPE trend was radically different during the second and third industrial revolutions, when science preceded technology. In the second industrial revolution, led by electricity in the late 19th century, Maxwell’s theory of electromagnetism was established before the large–scale deployment of electrical grids. In the third revolution, based on quantum mechanics and microelectronics, transistors, semiconductors, and microprocessors were direct applications of theoretical advances. During both revolutions, EPE did not collapse in the way observed during the coal–and–steam transition. Although energy consumption largely increased \cite{gentvilaite2014role}, productivity gains were large and EPE trends of countries were smoother and generally positive \cite{kanderinternational}. The contrast with the first revolution is striking: when technology follows a consolidated scientific framework, the energetic penalty is mitigated.

Looking at the twentieth century more broadly, both total energy consumption and EPE increased. Efficiency improvements were significant, but population and output grew even more. Per capita energy use increased, and when multiplied by population growth led to a strong rise in total world energy consumption~\cite{csereklyei2016energy,stern2018energy,stern2019energy}. Nevertheless, the capacity of advanced economies to generate more GDP per unit of energy also improved, producing an upward trend in EPE.

The lesson from history is consistent. Industrial revolutions disrupt the energy–growth relation. In the early phase, profitability dominates and efficiency is disregarded. EPE falls. In the mature phase, scientific knowledge, technological refinement, and institutional adaptation restore and enhance EPE. The pattern repeats: shock, decline, recovery.

This perspective justifies placing EPE at the centre of the present discussion.
As the first industrial revolution, the AI revolution resembles a pre-scientific industrial revolution, driven by technological breakthroughs but lacking a fully established scientific paradigm. The historical record suggests that a fall in EPE is a concrete risk. The difference today is that we enter this new phase with global awareness of climate constraints and sustainability~\cite{Morain_2024}, and with the possibility that AI itself could be used to optimize the energy–GDP relation~\cite{Cheng_Yu_2019}.

\section{Recent EPE trends by country clusters}

\begin{figure}
    \centering
    \includegraphics[width=11.5cm]{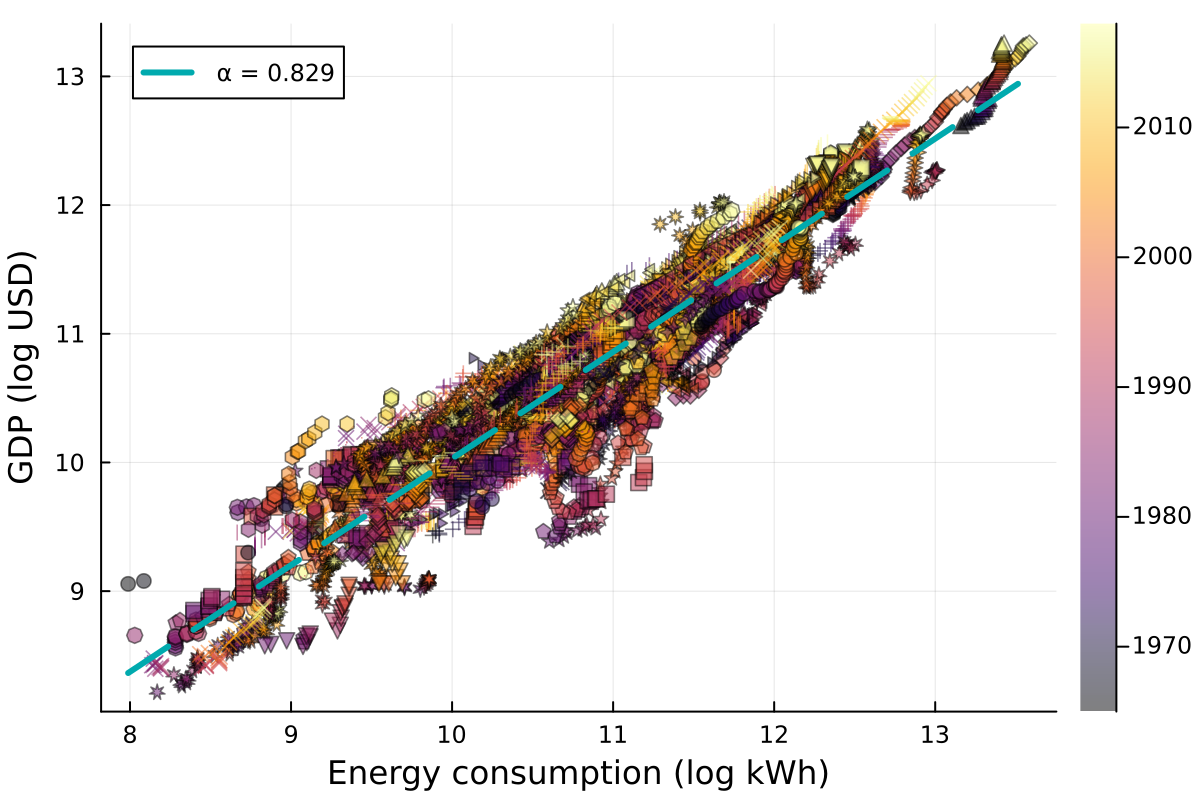}
    \caption{\textbf{Countries' GDP vs Energy consumption, over time.} Different years are represented by different colors as shown by the colorbar. Different countries have different markers. The measured Spearman coefficient value is $0.94$.}
    \label{smerlak}
\end{figure}

The data used for this study is obtained  from the following databases: the U.S. Energy Information Administration (2023) \cite{USEIA2024}; Energy Institute - Statistical Review of World Energy (2024) \cite{EnergyStats2024}; Bolt and van Zanden - Maddison Project Database 2023 \cite{MaddisonDatabase2024}  – with major processing by Our World in Data \cite{EnergyIntensity2024}. There, GDP measures are adjusted for inflation and differences in the cost of living between countries (\textit{Purchasing Power Parity}). Energy use refers to the use of primary energy before transformation to other end-use fuels, which is equal to indigenous production plus imports and stock changes, minus exports and fuels supplied to ships and aircraft engaged in international transport. The data used in this study was analysed using the Julia Programming software version 1.9.3. Figure \ref{smerlak} displays the log-log relationship between total energy consumption in kWh and total GDP in USD  from 1965 to 2018. An almost linear relation between GDP and energy consumption is found on the log-log scale. Furthermore, a very high Spearman correlation \cite{spearman1961proof} indicates that the two variables are strongly monotonically related.

We analyze EPE trends based on a country's classification into one of three macro groups: advanced, developing, or underdeveloped economies. To do so, we consider representative economies from each category. The division into these three clusters is performed following \cite{IMFRealGrowth2023, UNCTAD2023, DevelopingC_2024}. The ISO alpha-3 codes for the countries used in our analysis are as follows:
\begin{itemize}
    \item Advanced Economies: AUS, AUT, BEL, CAN, CYP, CZE, DNK, FIN, FRA, DEU, GRC, HKG, ISL, IRL, ISR, ITA, JPN, LUX, MLT, NLD, NZL, NOR, PRT, PRI, SGP, KOR, ESP, SWE, CHE, TWN, GBR, USA.
    \item Developing Economies: ALB, DZA, ARG, BHR, BRB, BOL, BWA, BRA, BGR, CMR, CPV, CHL, CHN, COL, COG, CRI, CIV, DMA, DOM, ECU, EGY, SLV, GNQ, SWZ, GAB, GHA, GTM, HND, HUN, IND, IDN, IRN, IRQ, JAM, JOR, KEN, KWT, LBN, LBY, MYS, MUS, MEX, MNG, MAR, NIC, NGA, OMN, PAK, PAN, PRY, PER, PHL, POL, QAT, ROU, LCA, SAU, SYC, ZAF, LKA, SYR, THA, TTO, TUN, TUR, URY, VEN, VNM, ZWE.
    \item Underdeveloped Economies: AFG, AGO, BGD, BEN, BFA, BDI, KHM, CAF, TCD, COM, COD, DJI, ETH, GMB, GIN, GNB, HTI, LAO, LSO, LBR, MDG, MWI, MLI, MRT, MOZ, MMR, NPL, NER, RWA, STP, SEN, SLE, TZA, TGO, UGA, YEM, ZMB.
\end{itemize}
These countries were selected based on the availability of data. Here, we are interested in two aggregate measures of EPE: one considers the total GDP of the countries in the cluster over the total energy,
\begin{equation}\label{eff_country}
    EPE_{C} = \frac{\sum_{i\in C} GDP_i}{\sum_{i\in C}E_i}
\end{equation}
while the other considers the weighted average by country's population:
\begin{equation}\label{eff_Pop}
    EPE_{C}^{(Pop)} = \frac{\sum_{i\in C} \frac{GDP_i}{E_i}Pop_i}{\sum_{i\in C}Pop_i}.
\end{equation}
Here $GDP_i, Pop_i$, and $E_i$ are respectively the gross domestic product, the population, and the energy consumption of country $i$, belonging to cluster $C$. The two measures \eqref{eff_country} and \eqref{eff_Pop} can generally be different. Equation \eqref{eff_country} is the ratio of the total GDP of all countries in cluster $C$ to the total energy consumption of all countries in $C$. It represents the average GDP generated per unit of energy used by the entire cluster $C$, giving more importance to countries that consume more energy. This gives an idea of how efficiently an entire cluster is converting energy into GDP, with a focus on total energy use. The second measure, equation \eqref{eff_Pop},  is the weighted mean of GDP per energy, where the weighting factor is the population of each country. Here, countries with larger populations have more influence on the overall average. For instance, highly populated countries with relatively low values of GDP and energy are more influential in $EPE_C^{(Pop)}$ than in $EPE_C$. If their EPE is high with respect to the other countries in the cluster, then $EPE_C^{(Pop)}>EPE_C$.
\begin{figure}[h!]%
\centering
\subfigure[Mean of the cluster (eq. \eqref{eff_country})]{%
\label{mean efficiency cluster}%
\includegraphics[height=2.85in]{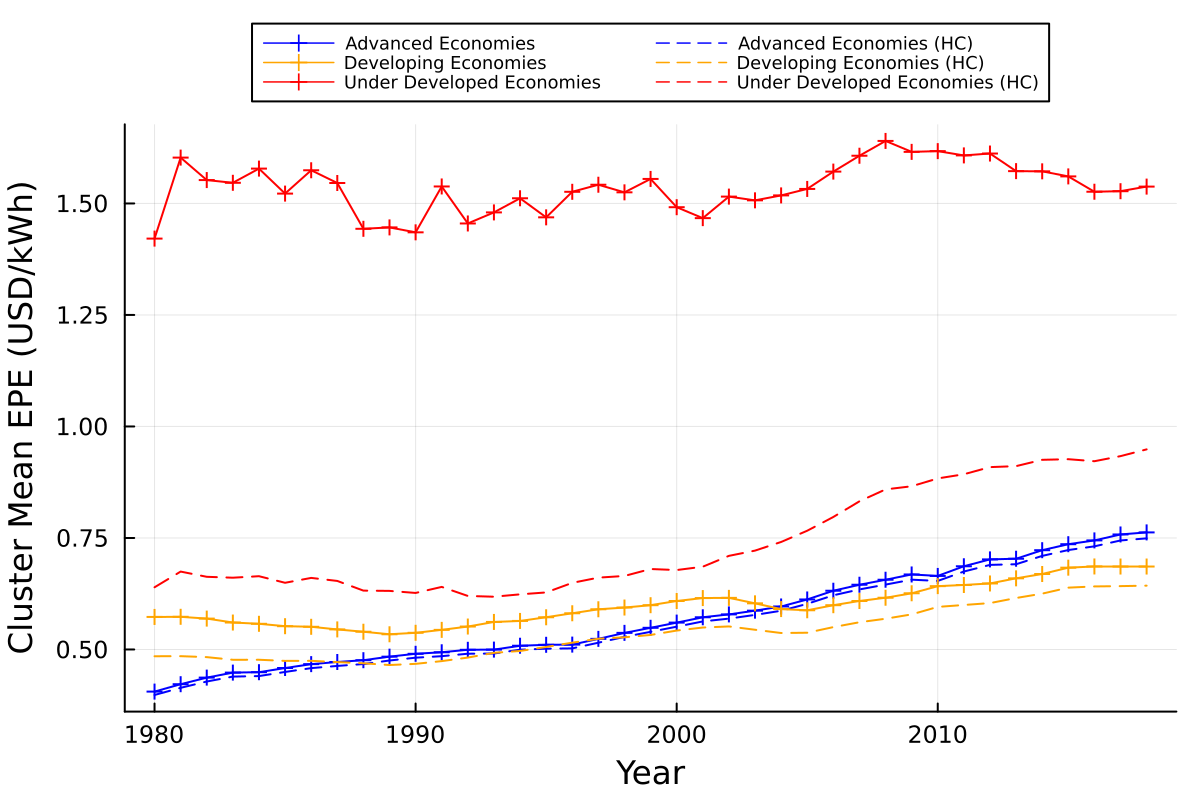}}\\
\subfigure[Mean of the cluster weighted by population (eq. \eqref{eff_Pop}) ]{%
\label{mean efficiency weighted by pop}%
\includegraphics[height=2.85in]{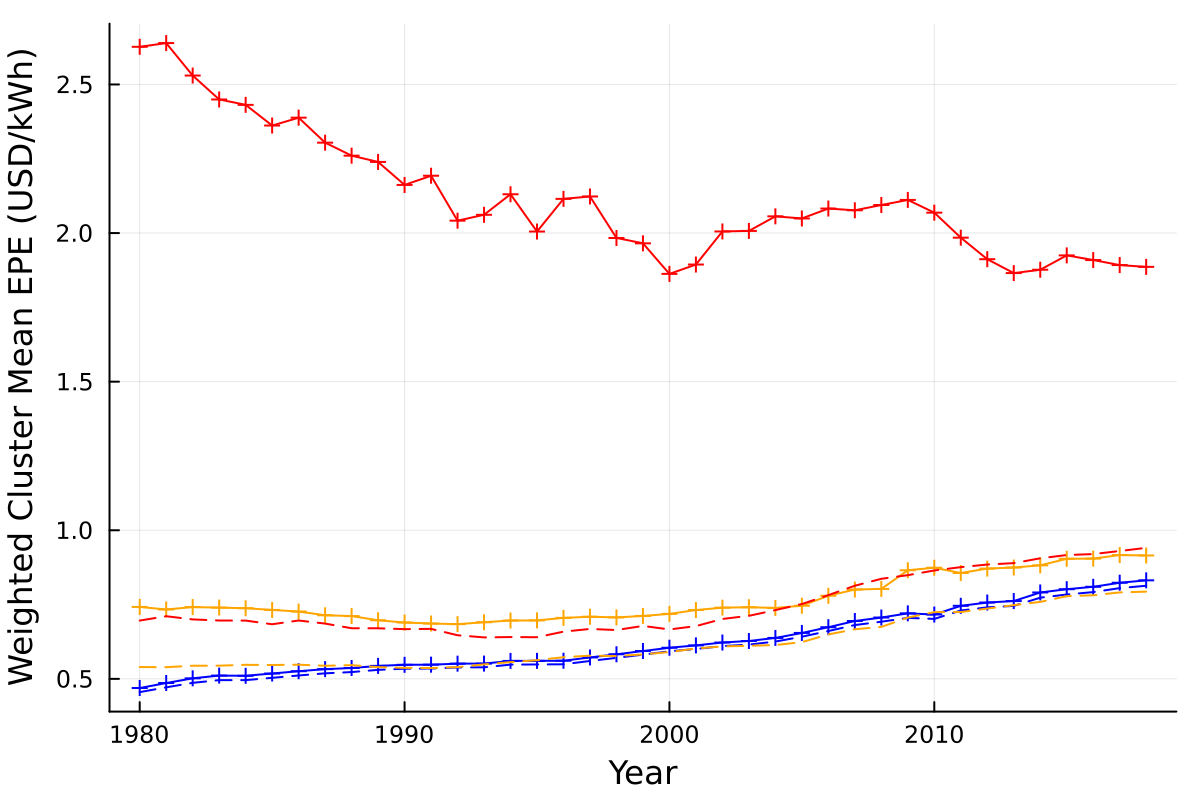}} 
\caption{\textbf{Mean EPE for different clusters of countries over time, without (solid lines) and with (dashed lines) human energy consumption.} The trend of the advanced economies is given by the blue curve, the developing economies by orange, and the underdeveloped economies by red.
}
\label{efficiency with and without HEC}%
\end{figure}

We find that the mean EPE of the advanced economies has increased monotonically in the last 40 years in both Figure \ref{mean efficiency cluster} and \ref{mean efficiency weighted by pop} almost linearly. It can be observed from Figure \ref{mean efficiency cluster} that around 2004 the mean EPE of the advanced economies surpassed the developing economies. The EPE of the latter exhibits a persistent growth only after around 2005.

Globally, world EPE grew monotonically in the last 40 years \cite{EnergyIntensity2024}, and the growth was driven by the improvement of the EPE of the advanced economies. Furthermore, we find that the underdeveloped economies have the highest values of EPE. The nearly constant trend of the mean EPE of the underdeveloped economies in Figure \ref{mean efficiency cluster} suggests that, overall, the cluster's ability to generate GDP relative to its energy consumption has remained stable over time. Nevertheless, in Figure \ref{mean efficiency weighted by pop}, we observe that when weighted by countries' populations, the average EPE is generally higher and exhibits a decreasing trend. That means that countries with relatively low values of energy consumed and GDP with respect to their population have typically higher values of EPE. If we assume that the GDP of the underdeveloped countries is produced mainly by human work, the data suggests that, in general, organized human work at a societal level requires less energy to generate the same amount of income as industrial machines. 

In order to account for the energy consumed by human labour, we modify the measure of energy consumption in a country by adding a term proportional to the country's population.  Considering a daily energy use per individual of $2500 \;\mbox{kcal}/\mbox{day} = 2.9$ $kWh$ \cite{owid-calorie-supply-sources}, we modify the measure of energy consumption as follows:
\begin{equation}\label{ene_pop_with human} E_i' = E_i+2.9\cdot 365\cdot Pop_i, 
\end{equation}
where $E_i'$ denotes the total energy accounting for human consumption, and $E_i$ is the energy consumed by machines. The value of 2.9 kWh per day is based on the world average energy consumed by a human being. We emphasize that this correction does not restrict the added term to the working-age population, as every individual, regardless of age, consumes energy for basic biological functions. The goal is to account for the total societal metabolic energy, not just the economically productive share. The impact of incorporating human energy consumption is illustrated by the dashed lines in Figure \ref{efficiency with and without HEC}, where one observes a significant decrease in the mean EPE for underdeveloped economies. The perturbation is more pronounced for underdeveloped and developing economies, whose GDP and energy consumption values are lower, while the EPE of advanced economies remains almost unchanged. Taking human energy consumption into account, we find that the EPE of the three clusters generally exhibits less dispersion. Finally, we note that these results are not highly sensitive to the specific measure chosen.

\section{AI as a pre-scientific engine: risks and opportunities for EPE}

Artificial intelligence is a general-purpose technology with economy–wide effects. Its diffusion is rapid, its impact heterogeneous across sectors, and its scientific foundations are still under construction. In this sense AI is a pre-scientific engine: the technology scales and monetizes before a consolidated theory of its limits, efficiencies, and system–level externalities is in place. Historically, pre–scientific phases of general–purpose technologies coincide with shocks to energy productivity. The relevant question is not whether AI will affect the Economic Productivity of Energy (EPE), but how and on what timeline.

\paragraph{Mechanism on the growth side.}
Macroeconomic assessments converge on a positive contribution of AI to labor productivity and GDP~\cite{acemoglu2019automation,autor2022new,czarnitzki2023artificial,behrens2020industry,bessen2019shocking,chui2023economic}, with adoption dynamics that are non–linear and delayed~\cite{brattberg2020europe,Mckinsey2019,hatzius2023potentially}. The expected channels are standard: task automation, decision support, quality augmentation, reallocation of labor toward higher–value activities, and creation of new products and services. If the energy required to deliver AI–enabled services grows more slowly than AI–driven output, EPE improves. If energy grows faster, EPE deteriorates. 

\paragraph{Mechanism on the energy side.}
Training frontier models requires concentrated bursts of computation. These bursts translate into multi–day or multi–week loads in the megawatt to gigawatt–hour range for large systems~\cite{de2023growing}. Inference, once models are deployed, adds a sustained, high–throughput load that scales with users, tokens, and latency constraints. Unlike conventional cloud workloads, which are more elastic and often I/O bound, modern AI workloads are compute bound and energy intensive, with non–negligible cooling and water footprints~\cite{Morain_2024}. Data centers follow the electricity mix of the grids they connect to. As deployments expand, location and grid composition become first–order determinants of the energy and carbon intensity of AI services. The energetic side of the EPE, related to AI, moves through hardware, software, and infrastructure choices.

\paragraph{AI impact on EPE: two possible scenarios.}
The current wave of AI resembles the early diffusion of steam power: profit opportunities appear before high efficiency is engineered and standardized, through theoretical understanding. Model size and training compute grew faster than algorithmic efficiency for several cycles. Supply chains respond by adding capacity: new data centers, accelerated computing clusters, dedicated interconnects, more cooling, and long–term power purchasing agreements. The initial equilibrium may be growth-first, as in the first industrial revolution. In such a scenario a temporary decline of EPE is plausible: output rises, but energy rises faster. Only when the technology is redesigned for efficiency, and AI technologies become well integrated in the economic and societal tissue, does the ratio improve. In an alternative scenario, pressures for sustainability delay AI adoption and diffusion until their energetic costs become manageable and their contribution to economic growth more consistent. In this case, EPE would avoid any decline and instead maintain or even strengthen its current positive trajectory.

\paragraph{Heterogeneity, risk and opportunity.} AI's impact on energy productivity will be heterogeneous across sectors and geographies. Advanced economies hold the greatest near-term potential for AI-driven GDP growth~\cite{Mckinsey2018}, and this may drive the future trend of EPE, as these are the countries that, with their high levels of GDP and energy consumption, mainly contribute to global EPE. On the other hand, emerging economies may adopt AI later and differently, with an efficiency path that depends on whether AI arrives alongside efficient infrastructure or through imported services. Risks include grid stress, siting issues and water constraints, and carbon-intensive deployments that could lock in high energy intensity, compounded by voluntary reporting that weakens optimization pressure. Conversely, significant opportunities exist for AI to boost system-wide energy productivity through grid optimization, forecasting, and smart controls, while transparency standards could redirect innovation toward quality-per-kWh rather than quality-per-request, enabling compounded efficiency gains across the economy.

\section{Policy implications and measurement agenda}

The following section integrates the preceding analysis into a broader policy perspective. It highlights why the Economic Productivity of Energy (EPE) should complement or substitute traditional indicators, how EPE can be monitored more effectively, and how it can inform policy responses to emerging technological shocks, most notably the impact of AI revolution on sustainability and economic growth.

\paragraph{Why EPE index.} While energy intensity is a well-established index, its inverse - EPE - remains largely underused. The distinction between the two is interpretative: whereas energy intensity measures how much energy is required to produce a given economic output, EPE expresses how much economic value can be generated from a fixed energy budget. In a world constrained by climate targets and finite resources, this is a more relevant metric, one that policy frameworks should explicitly acknowledge. Framing in terms of productivity rather than cost changes incentives and policy narratives. A fall in energy intensity can be interpreted as a marginal gain. A rise in EPE is seen as a direct productivity improvement.

\paragraph{Monitoring EPE: frequency and granularity.} The infrastructure for monitoring EPE is already in place: data on GDP and energy are already collected, and energy intensity calculated. The additional step is to publish the ratio systematically, with disaggregation by sector and by technology adoption. AI infrastructure scales rapidly: training runs and deployments produce shocks within months. EPE should therefore be reported at least quarterly for advanced economies and major emerging economies. Granularity matters as well. Aggregate ratios are informative but hide sectoral heterogeneity. Energy use in AI data centers, for example, is a small share of total consumption but grows fast. Sectoral EPE indicators can reveal whether efficiency gains in one area compensate for burdens in another. Overall, a frequent and granular EPE monitoring can detect signals of a potential EPE decrease.

\paragraph{Risk of misinterpretation.}
EPE must not be used as a simplistic performance score. High values in underdeveloped economies are an artifact of omitted biological energy. Low values in industrializing countries can reflect transitional dynamics rather than failure. Policy must read EPE historically and structurally, not mechanically. The lesson from past revolutions is that temporary declines are possible and sometimes unavoidable. The policy task is to shorten them and accelerate recovery. The indicator is diagnostic, not prescriptive.

\paragraph{Policy leverage.} EPE can serve as a policy lever across multiple scales. At the infrastructure level, incentives may be linked to efficiency benchmarks, with subsidies or permits granted only when EPE remains stable or improves. At the corporate level, disclosure standards and tax instruments can be calibrated to reflect productivity performance. At the societal level, public investment can prioritize research and deployment pathways that demonstrably enhance EPE. The framework is adaptive: rather than prescribing specific technologies, it delineates the outcomes policy should pursue.

\begin{tcolorbox}[title=\textbf{Policy Box: Monitoring EPE in the AI Transition},colback=white]
\begin{itemize}
  \item Establish EPE as an official indicator in sustainability reports. 
  \item Publish quarterly EPE dashboards, with sectoral detail and explicit tracking of AI-related energy use.
  \item Condition major AI infrastructure incentives on demonstrated EPE neutrality or improvement.
  \item Mandate standardized disclosure of AI training and inference energy, including cooling and water use, harmonized across jurisdictions.
  \item Fund cross–disciplinary research on the thermodynamics of computation and the macroeconomics of energy productivity.
  \item Promote research that connects the physics of computation with the economics of energy use.
\end{itemize}
\end{tcolorbox}
 
\paragraph{Summary.}The EPE is a compact ratio with broad policy relevance and represents a more appropriate indicator of efficiency in a world constrained by limited resources, compared to the more established energy intensity. Historical evidences show that the first industrial revolution, which has been pre-scientific in the sense that technological development preceded theoretical understanding, disrupted this measure, at least in its early stages. For successive, post-scientific, revolutions, the impact has been much softer or positive. The ongoing transition, driven by AI, is similarly pre-scientific and may follow a comparable trajectory, with a dip-and-rebound effect in EPE. This outcome is unsustainable and increasingly unacceptable today. Policymakers should therefore aim to monitor and actively steer this indicator, to detect early signals of the AI revolution’s impact on this crucial metric and to implement measures that mitigate potential efficiency losses while fostering a sustainable technological transition.

\section{Conclusions}
The Economic Productivity of Energy (EPE) provides a concise measure of how efficiently societies transform energy into economic value. Historical evidence from the first industrial revolution suggests that when technological diffusion precedes scientific understanding, EPE can temporarily decline before recovering through efficiency gains and societal adaptation. In later industrial revolutions, which unfolded on a more post-scientific basis, these transitions instead tended to produce smoother or even positive trends in EPE. This highlights a general risk: major technological revolutions can challenge the energetic foundations of growth when profitability advances faster than theoretical comprehension.

Artificial intelligence exemplifies a pre-scientific technology: we lack a consolidated theoretical understanding of its systemic efficiencies and limits. While AI has the potential to drive economic growth, it also consumes substantial energy. Whether its overall effect on EPE will be positive or negative remains uncertain, depending on how efficiency, transparency, and regulation evolve alongside innovation. Assessing the energetic dimension of AI’s contribution to growth is therefore crucial to avoid repeating past misalignments between technological progress and energetic sustainability.

Forecasts indicate that the impact of AI on EPE will likely be heterogeneous across countries, depending on their level of development. In this paper, we find that advanced economies have exhibited approximately linear growth in EPE for decades, in contrast to clusters of developing and underdeveloped countries. Thus, it should be noted that the impact of AI will occur on top of an already positive trajectory in these advanced economies.

Regular monitoring of EPE can make this balance visible. Systematic reporting of GDP per unit of energy, with sectoral detail and transparent accounting of AI-related energy use, would allow early detection of declining productivity and support timely policy responses. Embedding this indicator within sustainability frameworks would help ensure that technological progress continues to generate value within the boundaries of energetic and environmental constraints.

In this sense, tracking EPE serves both as a diagnostic and governance tool: it enables observation of whether the ongoing technological transformation enhances the efficiency of energy use, a necessary condition for sustainable growth in the decades ahead.

\section*{Acknowledgements}
PC thanks Fu Jun and Angelo Maria Petroni for the exchange of ideas that led to the investigations carried out in this paper. We thank Matteo Smerlak for his insightful discussions and important suggestions.  We are grateful to David Stern and Paolo Malanima for their valuable suggestions on an earlier version of the paper. We acknowledge the kind hospitality of the International Center for Theoretical Physics (Trieste), where this work was started. This project was supported by the EU H2020 ICT48 project Humane AI Net through grant N. 952026, the PRIN 2022 - code: J53D23003690006 and the Italian Extended Partnership PE01- FAIR Future Artificial Intelligence Research- Proposal code PE00000013 under the MUR National Recovery and Resilience Plan.

\nocite{Ang_2006,Bristol_de2024,Broughel_2024,Cheng_Yu_2019,Cuthbertson_2024,DevelopingC_2024,Google100Renewable,Halper2024,McQuate,MediumChatGPTConsumption,MicrosoftSus,Morain_2024,Murphy_2022,PWC,Petroni,Plumber2024,Reuters,StatistaGoogleConsumption,StatistaRenewable,WillWade,YannLeCun,amazon,gaida2023aspi,mehrara2007energy,patterson2022carbon,spearman1961proof}
\bibliographystyle{unsrt}
\bibliography{ref}

\end{document}